\documentclass[a4paper]{article}
\usepackage{Odyssey2020}
\usepackage{epsfig,amssymb,amsmath}
\usepackage{cite}
\ninept

\title{Using Multi-Resolution Feature Maps with Convolutional Neural Networks for Anti-Spoofing in ASV}

\name{Qiongqiong Wang, Kong Aik Lee, Takafumi Koshinaka}

\address{Biometrics Research Laboratories  \\
NEC Corporation, Japan \\
{\small \tt q-wang@nec.com} }
\begin{document}
\maketitle

\begin{abstract}
This paper presents a simple but effective method that uses multi-resolution feature maps with convolutional neural networks (CNNs)
for anti-spoofing in automatic speaker verification (ASV). 
The central idea is to alleviate the problem 
that the feature maps commonly used in anti-spoofing networks 
are insufficient for building discriminative representations of audio segments, as they are often extracted by a single-length sliding window. 
Resulting trade-offs between time and frequency resolutions restrict the information in single spectrograms. 
The proposed method improves both frequency resolution and time resolution by stacking multiple spectrograms 
that are extracted using different window lengths. 
These are fed into a convolutional neural network in the form of multiple channels, 
making it possible to extract more information from input signals while only marginally increasing computational costs.
The efficiency of the proposed method has been conformed on the
ASVspoof 2019 database. 
We show that the use of the proposed multi-resolution inputs consistently outperforms that of score fusion across different CNN architectures. Moreover, computational cost remains small.

\end{abstract}

\section{Introduction}

While Automatic speaker verification (ASV) offers flexible biometric authentication and has been increasingly employed in such telephone-based services as telephone banking, in forensics,  
at call centers, and in much mass-marketing of consumer products, 
its reliability depends on its resilience to intentional circumvention, i.e., spoofing,
as is true of any biometric technology \cite{Wu2015b}.

Attention to spoofing detection has increased significantly
with the dramatically increased use of biometric technology,
and effective anti-spoofing technology is essential for the commercial use of ASV in biometric authentication. 
There are four types of spoofing attacks w.r.t. to ASV:
impersonation, replay, text-to-speech speech synthesis \cite{Leon2012},
and voice conversion \cite{Wu2014}, among which replay is easiest to implement and the hardest to detect.


ASVspoof Challenges have been driving efforts on anti-spoofing measures \cite{Evans2013, Wu2015a,Delgado2018,asvspoof}. 
ASVspoof 2015 \cite{Wu2015a} focused on promoting
awareness and fostering solutions to spoofing attacks generated
from speech synthesis and voice conversion, while ASVspoof 2017 \cite{Delgado2018} 
focused on replay attacks. 
ASVspoof 2019 \cite{asvspoof} addressed both logical and physical access scenarios 
and further extended datasets in terms of spoofing technology, 
numbers of conditions and volumes of data.

The ASVspoof Challenges have resulted in significant findings.  
Constant-Q cepstral coefficient (CQCC) \cite{Todisco2016} features, 
which use constant-Q transforms (CQTs) rather than 
Fast Fourier Transforms (FFTs)  to process speech signals, 
perform better than ordinary Mel-frequency cepstral coefficients (MFCCs).
CQCC with Gaussian Mixture Models (GMMs) \cite{Reynolds1999} is now
a standard system used in spoofing detection for ASV. 
In \cite{Patel2015}, Cochlear filter cepstral coefficients (CFCCs) and
changes in instantaneous frequencies (CFCCIFs) have been proposed for training two simple GMM classifiers for the detection of  genuine and spoofing speech.

Recently,  the use of high time-frequency resolution features
has become a popular approach \cite{Suthokumar2018,Sahidullah2015,Patel2015}.
Higher accuracy has been achieved by directly 
using CQT spectrograms, 
from which CQCC features are extracted, 
together with deep neural networks (DNNs). 

The use of convolutional neural networks (CNNs) has been shown to perform much better than using GMM directly \cite{Sriskandaraja2018,Tom2017,Chen2017,Lavrentyeva2017}.
Light CNNs (LCNNs) with a max-feature-map (MFM) activation function \cite{Lavrentyeva2017,Wu2018}
extract significantly high-level embeddings from log-power spectrograms, 
which are obtained via CQT or FFT \cite{Sriskandaraja2018, Lavrentyeva2017}.  
When binary classes are well-separable in a high-level feature space, 
it is useful to employ simple two-class GMMs to obtain log-likelihood ratios (LLRs).

Feature map extraction is essential in speech processing tasks, including spoofing detection.  
Utilizing only one type of acoustic feature 
is insufficient for detecting global spoofing factors when facing unseen spoofing speech \cite{Li2019}. 
From a given audio segment, more than one acoustic feature map can often be extracted.
Different settings used in the extraction of feature maps will result in the obtaining of differing information. 
For example, FFT spectrograms \cite{Oppenheim1999}
extracted with different window lengths contain spectral information having resolutions that differ on higher and lower frequency bands,
and multiple FFT spectrograms might be obtained by 
using different window lengths in the extractions.  
A longer window length will lead to higher resolution in terms of frequency 
and lower in terms of time.
Conversely, a shorter window length will result in higher resolution in terms of time but lower in terms of frequency.
It may be difficult, if not impossible, to determine the one type of acoustic feature maps that will be best for  spoofing  detection,
particularly when different neural network structures are used.
The trade-off between time and frequency resolution makes it difficult to extract sufficient information with one FFT spectrogram alone.
This is also true for other acoustic feature maps, such as CQT and  
MFCC, for which using just one extraction configuration will limit the amount of information obtained,  
and  the use of multiple acoustic feature maps is needed to alleviate the problem.
In general, different feature maps compliment one another and  
help obtain information that is more highly discriminative.

Feature fusion and score fusion 
make use of the multiple feature maps. 
Feature fusion, a.k.a. early fusion, includes feature map concatenation along a single dimension, 
such as a time or frequency dimension. 
Linear interpolation is an alternative feature fusion method. 
Score fusion, a.k.a. late fusion, can be used to fuse scores produced from systems using individual feature maps.
However, score fusion can be computationally costly since it needs to train neural networks for multiple times. 
In addition, when scores are fused, 
weights need to be determined in advance,
which may not be easy to do 
since optimum weights often differ depending on the application being used.

We propose a  simple  but  effective  method  
that uses multi-resolution feature maps  for  anti-spoofing  in deep  neural networks.
It stacks multiple feature maps of the same dimensionality into a three-dimension input for deep neural networks.
Our aim is to alleviate the problem that the feature maps commonly used in anti-spoofing networks are insufficient for building discriminative representations of audio segments,
as they are often extracted by a fixed-length sliding window. 
With our proposed method, multi-resolution feature maps are fed into neural networks in the form of multiple channels.   
This makes it possible to extract more information from input signals with relatively little
computational cost.
The efficiency of the proposed method has been confirmed on the ASVspoof 2019 database \cite{asvspoof} .   We show that the used of the proposed multi-resolution  inputs  consistently  
outperforms that of score fusion across different DNN architectures.   
Moreover,  the  computational  cost remains small.

The remainder of this paper is organized as follows: 
Section 2 describes feature extraction commonly used in spoofing detection and the proposed multi-resolution feature maps combined with CNNs.
Section 3 describes our experimental setup, results, and analyses; 
and Section 4 summarizes our work.

\section{Proposed System}
\subsection{Feature Extraction}
In this paper we use 
Fast Fourier Transforms (FFTs) \cite{Sriskandaraja2018} for spoofing detection.

\subsubsection{Fast Fourier Transforms (FFTs) }
An FFT performs a Fourier Transform on a short segment that has been extracted from a longer data record upon its multiplication with a suitable window function. A sliding window is applied repetitively in order to analyze the local frequency content of the longer data record as a function of time \cite{Oppenheim1999}. 
The FFT is essentially a filter bank. The $Q$ factor is a measure of the selectivity of each filter and is defined as the ratio between the center frequency $f_k$ and the bandwidth $\delta$f:

\begin{equation}
Q = \frac{f_k}{\delta f}
\label{eq1}
\end{equation}
In the FFT, the bandwidth of each filter is constant and related to the window function. The Q factor thus increases when moving from low to high frequencies since the absolute bandwidth $f$ is identical for all filters. This is in contrast to the human perception system, which is known to approximate a constant Q factor between 500Hz and 20kHz \cite{Moore2003}. 


\subsubsection{Feature Engineering}
\begin{figure}[t]
\includegraphics[width=\columnwidth]{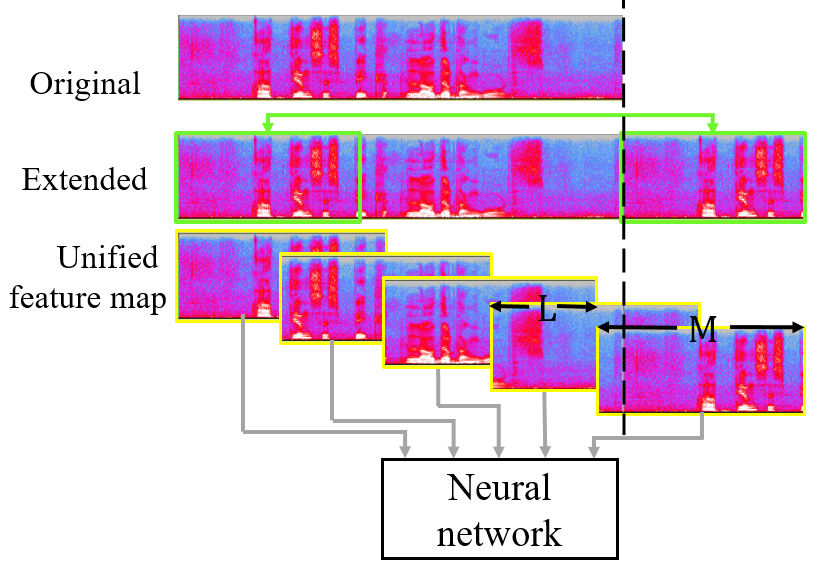}
\caption{{\it Illustration of the Unified Feature Map approach. Each feature map contains M frames, and the overlap between two feature maps consists of L frames.}}
\label{fig:feature_eng}
\end{figure}

We have followed the work in \cite{Lai2019a} in creating a unified feature map for use as input to CNN models. 

Since the lengths of evaluation utterances are usually not known beforehand, 
all utterances are extended in training and evaluation datasets to the minimal multiple of $M$ frames by repeating the audio segments, as illustrated in Figure~\ref{fig:feature_eng}. 
The extended feature map is then broken down into segments
of $M$ frames. 
The segments have an overlap of $L$ frames.

\subsection{Multi-Resolution Feature Maps for CNNs}

\begin{figure}[t]
\centering
\includegraphics[scale=0.7]{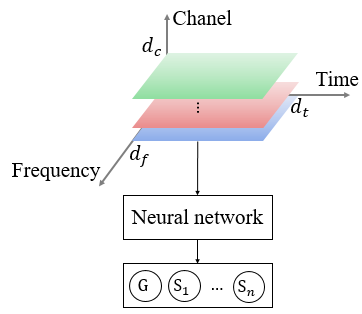}
\caption{{\it Illustration of the proposed deep learning using multi-resolution feature map input. 
Output node $``G"$ represents $``$genuine class$"$.
$``S_i"$ represents the $i^{th}$ spoofing class.}}
\label{fig:nn}
\end{figure}

Feature map extraction is essential in speech processing tasks,  including  spoofing  detection.   
Utilizing  only  one kind  of  acoustic  feature  is  insufficient  for  capturing  global spoofing factors  when  facing  unseen  spoofing  speech \cite{Li2019}.
Multiple feature maps for a single audio are often available.
For example, multiple FFT spectrograms can be extracted from a single audio segment by using different window lengths.
However, it is difficult, if not impossible, to determine which of the feature maps would be best for spoofing detection. 
As we have previously noted, a longer window length leads to an FFT spectrogram with higher resolution in terms of frequency and lower resolution in terms of time, 
while the converse is true for a shorter window.
The trade-off between time and frequency resolution makes it difficult to extract sufficient information with a single FFT,  
as is also true for such other feature maps as CQT and MFCC, 
while multiple feature maps can be used compliment one another. 

We propose the use of multi-resolution feature maps, which consist of  a stack  of multiple feature maps of the same dimensionality, into three-dimension inputs for convolutional neural networks (CNNs), as is shown in Figure ~\ref{fig:nn}. 
The modification to network is simple.
We need to change only the configuration $($input$\times$output$)$ of the first convolutional layer from
$(1\times{c_1})$ to $(n_c \times c_1)$,
where $n_c$ is the number of feature maps in the input of CNNs and
$c_1$ is the number of output dimensions of the first convolutional layer. 
Thus, the increase in the number of parameters will be small, as the first convolutional layer generally contains a few hundred output nodes.
Such increments are negligible in comparison to the total number of  neural network parameters, which is usually on the order of millions.

\begin{table}[t]
\caption{ {\it Model parameters of ResNet18 and SENet50. BN stands for a bottleneck residual block. Basic and Bottleneck residual blocks are described in the original ResNet \cite{He2016}}}
\vspace{2mm}
\centerline{
\begin{tabular}{c|l  |c| c| c| c    }
\hline
Model     & Config.      & Block1 & Block2 & Block3 & Block4\\
\hline  \hline
          & unit type    &  Basic &  Basic &  Basic &  Basic \\
ResNet18  & num. of unit &  2     &   2    &    2   &   2 \\ 
          & channels     & 16     & 32    &64     &128 \\
\hline
          & unit type    & BN     & BN    & BN    & BN  \\
SENet50   & num. of unit & 3      & 4     & 6     & 3   \\
          & channels     & 16     & 32    &64     &128  \\
\hline
\end{tabular}}
\label{tab:resnet}
\end{table}

\begin{table}[t]
\caption{\label{table2} {\it LCNN structure. MFM stands for max-Feature-map activation.}}
\vspace{2mm}
\centerline{
\begin{tabular}{l|c|c| c    }
\hline
Type      & Filter   & Stride & Channel \\
\hline  \hline
Conv1     & $5\times5$      & $1\times1$    & 32 \\
MFM1      & -        & -      & 16\\
\hline
MaxPool1  & $2\times2$      & $2\times3$    & 16\\
\hline
Conv2a    & $1\times1$      & $1\times1$    & 32 \\
MFM2a     & -        & -      & 16\\
Conv2b    & $3\times3$      & $1\times1$    & 48 \\
MFM2b     & -        & -      & 24\\
\hline
MaxPool2  & $2\times2$      & $2\times3$    & 24\\
\hline
Conv3a    & $1\times1$      & $1\times1$    & 48 \\
MFM3a     & -        & -      & 24\\
Conv3b    & $3\times3$      & $1\times1$    & 64 \\
MFM3b     & -        & -      & 32\\
\hline
MaxPool3  & $2\times2$      & $2\times3$    & 32\\
\hline
Conv4a    & $1\times1$      & $1\times1$    & 64 \\
MFM4a     & -        & -      & 32\\
Conv4b    & $3\times3$      & $1\times1$    & 32 \\
MFM4b     & -        & -      & 16\\
\hline
MaxPool4  & $2\times2$      & $2\times3$    & 16\\
\hline
Conv5a    & $1\times1$      & $1\times1$    & 32 \\
MFM5a     & -        & -      & 16\\
Conv5b    & $3\times3$      & $1\times1$    & 32 \\
MFM5b     & -        & -      & 16\\
\hline
MaxPool5  & $2\times2$      & $2\times3$    & 16\\
\hline
\hline
FC6       & \multicolumn{3}{|l}{Output:  $64\times2$}   \\
MFM6      &  \multicolumn{3}{|l}{Output:  64}   \\
FC7       &  \multicolumn{3}{|l}{Output:  10}  \\
\hline
\end{tabular}}
\label{tab:lcnn}
\end{table}

\section{Experiments}

\subsection{Experimental Settings}
The experimental data used in this study
was the Physical Access (PA) subset of ASVspoof 2019 Challenge \cite{asvspoof}.
It contained $48,600$ spoofed and $5,400$ bonafide utterances in the training partition, 
as well as $24,300$ spoofed and $5,400$ bonafide utterances in the development partition. 
The spoofed utterances were recorded under $27$ different acoustic configurations and $9$ replay configurations. 

To extract acoustic features (here, FFT spectrograms),  
we used window lengths of $18 ms$, $25 ms$, and $30 ms$. 
The FFT spectrogram dimension was $257\times400$.
All FFT spectrograms were extracted using the Kaldi speech toolkit \cite{kaldi}.

The length of segments was set to $M=400$, 
and the overlap between two segments was set to $L=200$. 
In training, each segment was counted equally in loss calculations. 
This may not be the best way for optimization since longer utterances had higher weights in optimization. 
In the future, all utterances should ideally be treated equally. 
In our evaluations, the score for each utterance was calculated by averaging DNN outputs over all segments of the utterance. 

\begin{figure}[t]
\centering
\includegraphics[width=\columnwidth]{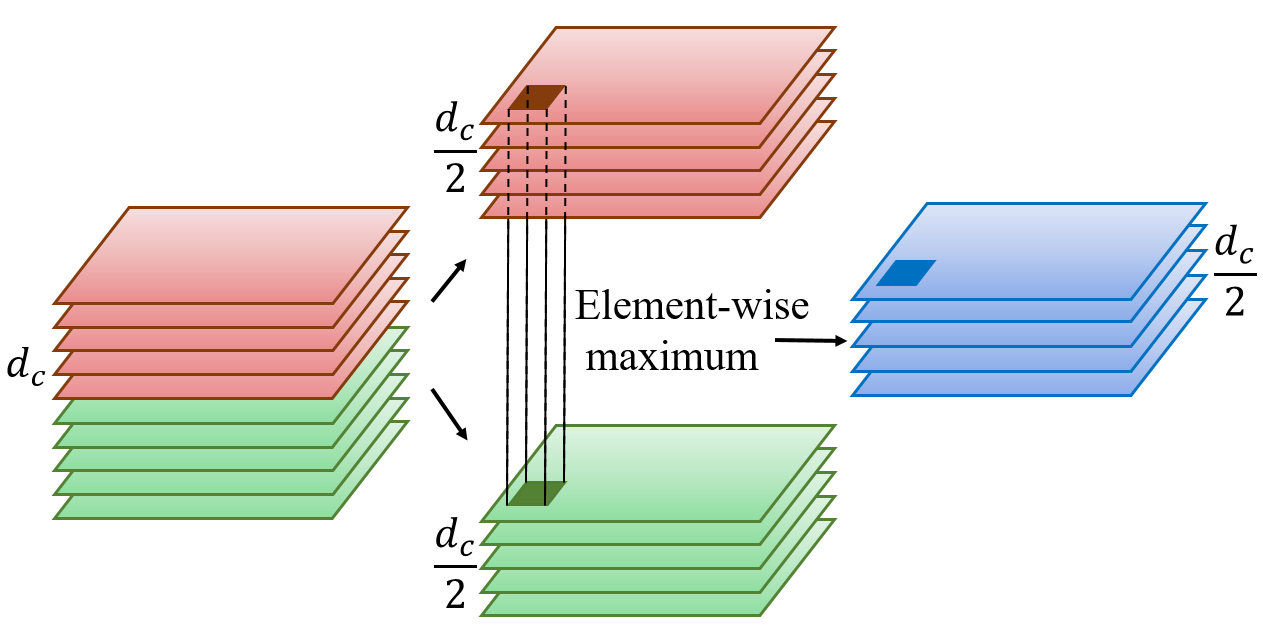}
\caption{{\it MFM for convolutional layers.}}
\label{fig:mfm}
\end{figure}

Experiments were carried out using the following three CNN variants: 
(1) ResNet18 \cite{Lai2019b,He2016};
(2) SENet50 (Squeeze-Excitation Network) \cite{Lai2019b}; and 
(3) Light CNN (LCNN) \cite{Lavrentyeva2017}.
The model parameters and architectures of ResNet18 and SENet50
are shown in Table~\ref{tab:resnet}. 
The basic and bottleneck residual blocks are described in the original ResNet paper \cite{He2016}.
LCNN is a kind of CNN with Max-Feature-MAP (MFM) activation.
Its architecture and model parameters are shown in Table~\ref{tab:lcnn}. 
The use of MFM allowed us to reduce CNN channels by half, 
which explains the use of the term Light CNN (see Figure ~\ref{fig:mfm}). 
For details, please see \cite{Lavrentyeva2017,Wu2018}. 
ResNet18 and SENet50 have been shown in \cite{Lai2019b} to be effective.
LCNN worked the best in the ASVspoof 2017 replay detection \cite{Lavrentyeva2017} and was ranked highly in the ASVspoof 2019 challenge. 
That is why we chose these three neural networks to evaluate our proposed method.
All three were trained using 64 segments in each mini-batch.

In network optimization,
we followed the optimization schemes described in \cite{Lai2019b} for training all three of the DNN models described above.
The output layer had 10 nodes representing a single bonafide condition and 9 spoofing conditions (different environments and methods of attack). 
For computing Equal Error Rates (EERs) during the evaluation stage, we took the
log-probability of the bonafide class as the score for a given utterance.

The DNN models were optimized using an Adam optimizer with $\beta_1 =
0.9$, $\beta_2 = 0.98$, and a weight decay of $10^{-4}$. 
The learning rate scheduler increased the learning rate linearly for the first 1000
warm-up steps and then decreased it proportionally to the inverse square root of the step number \cite{Vaswani2017}. Finally, after every
training epoch, we selected the best model on the basis of the EER of the development set.

\subsection{Results and Analysis}
We first compared spoofing detection EERs when using  single feature maps of different resolutions in ResNet18, SENet50, and LCNN, as shown in
Table~\ref{tab:single}. 
For different neural network architectures, 
the respective best performances were obtained with different feature maps. 
The FFT spectrograms extracted with $25ms$ and $30ms$ showed similar results in ResNet18, and significantly better than those with 18ms. 
For SENet50, however,  FFT spectrograms  with $18ms$ and $25ms$ gave similar results, while those with $30ms$ were best. 
For LCNN, the FFT spectrograms of $25ms$ gave the best performance,
i.e., there may not be one single optimal FFT configuration for differing neural network structures.

\begin{table}[t]
\caption{ {\it EER ($\%$) of ASVspoof 2019 PA development and evaluation set using single feature maps with ResNet18, SENet50 and LCNN.}}
\vspace{2mm}
\centerline{
\begin{tabular}{|l|l|r r|}
\hline
Network      & Feature map      & Dev  & Eval  \\
\hline  \hline
             & 18ms             & 6.44 & 6.36 \\
ResNet18     & 25ms             & 4.98 & 5.26 \\
             & 30ms             & 5.33 & 4.98 \\
\hline
             & 18ms             & 4.17 & 4.49 \\
SENet50        & 25ms             & 4.24 & 4.46 \\
             & 30ms             & 3.39 & 3.22 \\
\hline
             & 18ms             & 10.19 & 10.69 \\
LCNN         & 25ms             & 7.81  & 8.41 \\
             & 30ms             & 8.70 & 8.68  \\
\hline
\end{tabular}}
\label{tab:single}
\end{table}

Next, we applied the proposed multi-resolution feature maps as input to the CNNs. 
Results are shown in Table~\ref{tab:multi}. 
In ResNet18, 2-resolution input showed, respectively,
$22.7\%$ to $31.9\%$ and $12.2\%$ to $21.5\%$ lower EER in the development and evaluation sets as
compared to the better of the two single-feature systems.
3-resolution input showed the best performance: $52.0\%$ and $38.4\%$ lower EER as compared to the best of the three single-feature systems.

For SENet50, 2-resolution input showed, respectively, 
$42.2\%$ to $48.0\%$ and $25.8\%$ to $37.0\%$ lower EER in the development and evaluation sets as
compared to the better of the two single-feature systems.
3-resolution input showed the best performance: $57.5\%$ and $45.3\%$lower EER as compared with the best of the three single-feature systems.

For LCNN, 2-resolution input of FFT spectrograms extracted using $18ms$ and $30ms$ sliding windows achieved the best performance in both the development and the evaluation sets, 
for which EER was lower by $42.8\%$ and $22.9\%$, respectively. 

When multiple feature map inputs are available, it is a straight-forward process to feed each feature map to a CNN and fuse the resulting scores at the late stage, i.e.,  to employ score fusion.
Table~\ref{tab:multi} shows  a performance-comparison between 
our proposed method and score fusion.
Score fusion also resulted in better performance than that of the single-feature systems in Table~\ref{tab:single}.
The proposed method was significantly better than score fusion in all cases in all the three networks, and particularly so in ResNet18 and SENet50. 
It not only showed significantly better spoofing detection accuracy but also offered computational cost nearly equivalent to that of the original neural networks.

\begin{table}[t]
\caption{ {\it EER ($\%$) of ASVspoof 2019 PA development and evaluation set using multi-resolution feature maps and score fusions(conventional method) with optimum weights.}}
\vspace{2mm}
\centerline{
\begin{tabular}{|l|l|c c| c c|}
\hline
Network      & Feature map      &\multicolumn{2}{c|}{Proposed}  & \multicolumn{2}{c|}{Score fusion} \\
             &                  & Dev & Eval     & Dev & Eval \\
\hline  \hline
             & 18$||$25ms       & 3.85 & 4.62    & 4.89 & 5.06\\
ResNet18     & 18$||$30ms       & 3.63 & 3.91    & 4.94 & 4.61\\
             & 25$||$30ms       & 3.48 & 3.94    & 4.54 & 4.56\\
             & 18$||$25$||$30ms  & 2.39 & 3.07    & 4.56 & 4.47    \\
\hline
\hline
             & 18$||$25ms       & 2.17 & 2.90    & 3.36 & 3.61\\
SENet50        & 18$||$30ms       & 1.96 & 2.03    & 2.94 & 2.80\\
             & 25$||$30ms       & 1.88 & 2.39    & 3.13 & 3.04\\
             &18$||$25$||$30ms  & 1.44 & 1.76    & 2.87 &  2.82   \\
\hline
\hline
             & 18$||$25ms       &  6.20 & 7.98   & 7.74 & 8.26 \\
LCNN         & 18$||$30ms       &  4.98 & 6.69   & 8.36 & 8.31 \\
             & 25$||$30ms       & 5.72 & 7.50    & 7.63 &7.88\\
             &18$||$25$||$30ms  & 6.17 & 7.36       &7.65  &7.94\\
\hline
\end{tabular}}
\label{tab:multi}
\end{table}


\begin{table}[t]
\caption{ {\it Comparison of parameter numbers in systems using single and multi-feature maps and score fusion.}}
\vspace{2mm}
\centerline{
\begin{tabular}{|l|l|r|}
\hline
Network      & System                   & Parameter Num.  \\
\hline  \hline
             & 1 map                    & 701,808    \\
ResNet18     & 2 maps: proposed         & +784       \\
             & 2 maps: score fusion     & +701,808   \\
             & 3 maps: proposed         & +1,568     \\
             & 3 maps: score fusion     & +1,403,616 \\
\hline
             & 1 map                     & 1,094,640 \\
SENet50        & 2 maps: proposed         & +784      \\
             & 2 maps: score fusion     & +1,403,616\\ 
             & 3 maps: proposed         & +1,568    \\
             & 3 maps: score fusion     & +2,189,280\\
\hline
             & 1 map                    &  73,504    \\
LCNN         & 2 maps: proposed         &  +800     \\
             & 2 maps: score fusion     &  +73,504  \\ 
             & 3 maps: proposed         &  +1,600   \\
             & 3 maps: score fusion     &  +147,008 \\
\hline
\end{tabular}}
\label{tab:para}
\end{table}
As shown in Table~\ref{tab:para}, using the proposed 2-resolution feature maps only resulted in a parameter-number increase less than $0.12\%$, 
while the increase with use of the best 3-resolution feature maps was roughly  $0.22\%$.
Score fusion methods, as is well known, 
train two or more systems and fuse scores in the score level. 
This did not improve the performance significantly in our experiments, but it doubled or tripled the number of parameters, 
which means that our proposed method would be much more helpful in practical use.

\section{Summary}
This paper has presented a simple but effective method that uses multi-resolution inputs 
with convolutional neural networks  
for spoofing detection in ASV.  
Our aim is to alleviate the problem
that the feature maps commonly used in anti-spoofing networks 
are likely to be insufficient for building discriminative representations of audio segments, as they are often extracted by fixed-length windows. 
With the proposed method, multi-resolution feature maps,
which consist of a stack of multiple spectrograms, 
are fed into CNNs in the form of a multi-channel input, resulting in automatic selection of optimal resolutions.
The effectiveness of our proposed method has been confirmed on
the ASVspoof 2019 Physical Access (PA) database with ResNet18, SENet50, and Light CNN.
Experimental results show that use of 2-resolution feature maps results in EER lower by, respectively, nearly $21.5\%$ and $37.0\%$ with ResNet18 and SENet50 for the evaluation set, 
with only a $0.12\%$ increase in the number of parameters. 
It also achieved the best performance in LCNN, resulting in an EER reduction of $22.9\%$. 
Use of 3-resolution feature maps showed  the  best  performance for ResNet18 and SENet. It resulted in EER reductions of,  respectively, $38.4\%$ and $45.3\%$ for the same datasets. 
In future work, we intend to add an attention mechanism that makes better use of multi-resolution feature maps. Other feature extractors, such as CQT, CQCC, and MFCC, are to be examined as well.

\bibliographystyle{IEEEbib}
\bibliography{Odyssey2020_BibEntries}

%

\end{document}